# A Coverage Monitoring algorithm based on Learning Automata for Wireless Sensor Networks

Habib Mostafaei[1], Mehdi Esnaashari[2] and Mohammad Reza Meybodi[3]

[1] Department of Computer Engineering, Urmia Branch,
Islamic Azad University, Urmia, Iran
h.mostafaei@iaurmia.ac.ir

[2] Computer Engineering and Information Technology Department,
Amirkabir University of Technology Tehran, Tehran, Iran
esnaashari@aut.ac.ir

[3] Department of Computer Engineering and Information Technology,
Amirkabir University of Technology Tehran, Tehran, Iran

mmeybodi@aut.ac.ir

To cover a set of targets with known locations within an area with limited or prohibited ground access using a wireless sensor network, one approach is to deploy the sensors remotely, from an aircraft. In this approach, the lack of precise sensor placement is compensated by redundant deployment of sensor nodes. This redundancy can also be used for extending the lifetime of the network, if a proper scheduling mechanism is available for scheduling the active and sleep times of sensor nodes in such a way that each node is in active mode only if it is required to. In this paper, we propose an efficient scheduling method based on learning automata and we called it LAML, in which each node is equipped with a learning automaton, which helps the node to select its proper state (active or sleep), at any given time. To study the performance of the proposed method, computer simulations are conducted. Results of these simulations show that the proposed scheduling method can better prolong the lifetime of the network in comparison to similar existing method.

## 1. INTRODUCTION

Wireless sensor networks (WSNs) have been studied extensively in recent decade. They can be used in vast variety applications such as national security, surveillance, health care and environmental monitoring, to mention a few. Sensor nodes are small devices that can sense some phenomenon in the environment, process and save monitored data, and send data to a central node called the base station [1].

In WSNs, one of the most important design challenges is to increase network lifetime. This is especially critical when battery change is not applicable. In recent years, most research has been done on the efficient usage of battery resources to prolong the network lifetime. One of the common methods to improve lifetime is the node activity scheduling.

Node activity scheduling can be performed efficiently when sensor nodes are scattered redundantly to monitor a fixed placed list of targets. Every scheduling method must work around different performance requirements, for instance, routing connectivity, network coverage, redundancy requirement, etc. In this paper, we focus on target coverage problem and we assume that radio range of each node is enough large to maintain routing connectivity. In this case, each target in the network is covered by more than one sensor node, and hence, redundant sensor nodes can be put into sleep state to save their batteries, without affecting the overall coverage of the network. In addition coverage requirements, we wish to organize nodes in such way to prolong networks` lifetime.

Maximum set covers and maximum lifetime are two different problems in wireless sensor networks. In maximum set covers problem, every scheduling method try to schedule sensor nodes into set covers to increase network lifetime as each set cover can monitor all targets in network. In the past, most of the research focus was on dividing the sensor nodes into a number of disjoint subsets and at any given time, only one of the subsets is active to monitor the scattered targets [2,4,25]. The problem of how to find these disjoint subsets is referred to as disjoint set cover problem. The main objective on this type of problem was that how can we extend network`s lifetime.

In this paper, instead of dividing the sensor nodes into disjoint subsets, we introduce a learning automata based method for scheduling the active times of the sensor nodes without significantly affecting the network coverage. In this method, each node is equipped with a learning automaton and the learning automaton of each node helps the node to select its proper state (active or sleep) at any time during the operation of the network.

The rest of the paper is organized as follows. In Section 2, we present related works in the field of energy efficiency target coverage problem. Section 3 briefly describes the target coverage problem. Learning automata as a basic learning strategy used in the proposed method will be discussed in Section 4. In Section 5, the proposed method is presented. Section 6 presents the simulation results and Section 7 concludes the paper.

## 2. RELATED WORKS

Coverage problem has different definitions and specifications according to the recent researches in the wireless sensor networks. Zhu et al [17] provided a good survey on various coverage and connectivity issues in wireless sensor networks. Coverage problem mainly can be classified into three types: target (point) coverage, area coverage, and barrier coverage. The objective of point coverage problem is to cover a set of stationary or moving points. Scheduling sensor nodes into cover set is mostly used in different approach is used to solve this problem. In [24], the authors model the problem as a maximum cover tree problem and show that it is an NP-complete problem. They propose heuristic approximation algorithms to increase the lifetime of the network. In [23] authors proposed a cellular learning automaton based algorithm to monitor moving targets in networks. The main objective of area coverage problem  monitor the whole area of the network with respect to different performance criteria such as coverage ratio, minimum number of sensors providing desired coverage level during the maximum lifetime of the network. The node sleeping scheduling algorithms mostly are used to maximize network`s lifetime. In [18-19] authors proposed a learning automata based algorithm to monitor an area in wireless sensor networks. They used from learning automata as a method to select best sensor nodes among nodes` neighbors to monitor an area. Barrier coverage can be considered as the coverage with the goal of minimizing the probability of undetected penetration through the barrier (sensor network). This type of coverage problem needs less number of sensors than full coverage problem.

 In the target (point) coverage problem, the objective is to cover a set of disjoint fixed or moving targets. In the area coverage problem, the objective is to cover the area field of the network. Finally, in the barrier coverage problem, the main objective is to detect penetrated intrusion into the network.

In Cardei and Du [2] considered the target coverage problem, they proposed a centralized subset-based method which divides the sensor nodes into subsets, each can individually cover the entire targets. Their objective was to maximize the number of subsets and refer to the problem as

44

maximum set cover problem. They did not pose any limitations on the size of the network. Cardei and Wu in [4] proposed two Greedy heuristics for finding the maximum number of subsets, each capable of covering the entire targets. They also proved that the maximum set cover problem is NP-complete. In [12], Slijepcevic and Potkonjak addressed the area coverage problem where the area is modeled as a collection of fields and every field can be covered by the same subset of nodes in the networks.

In [18] authors present a survey in the field of coverage and connectivity problem. They reviewed evaluations of algorithms in the field of coverage and connectivity and also, they added additional metrics to evaluate the performance of methods that have presented. Maggie and Xuan [19] proposed two linear programming based algorithms for maximizing the lifetime of target coverage in wireless sensor networks. They showed the maximum lifetime problem is NP-complete. In [20], authors addressed multiple target coverage in wireless sensor networks and proposed two heuristic algorithms to prolong the network lifetime. Their algorithms compute maximum number of joint subsets for target coverage and they used the same approach in [2] to compute the lifetime of their algorithm.

In [22], authors presented a hybrid approximation approach for complete minimum-cost target coverage problem in wireless sensor networks. They used combination of LP-rounding and set cover selection method to propose their method. Slijepcevic and Potkonjak proposed column generation based algorithm to find near optimal solution for treatment target coverage in wireless sensor networks in [23]. They offered an approach that can guarantee at least $(1-\varepsilon)$ of optimal network lifetime.

In [21], authors proposed a distributed scheduling algorithm for special target coverage problem that called partial target coverage. In this problem, 100 percent target coverage is not required. They used residual energy level of each node and neighbors information as a feedback to propose their algorithm.

Another type of target coverage is called Connected Target Coverage (CTC) problem. In this problem, the objective is that monitor all deployed targets in network which each selected sensor nodes should connected to each other and sink node in network. Zhao and Gurusamy [24] considered connected target coverage problem in wireless sensor networks for special state in which each scheduled sensor node in network can communicate with each other and sink node directly or through multihop communication in network. They modeled this problem as maximum cover tree problem and proposed a greedy method to solve this problem. In [25] authors proposed an efficient method to guarantee coverage and connectivity in wireless sensor networks. They used a different deployment method to guarantees coverage and preserves connectivity. Also, in [26] another type of target coverage that called connected cover set is introduced. In this case, each subset selected sensor node can communicate with any other sensor node directly or via multihop communication in network.

Authors in [29] consider a sensor covers targets with users' satisfied probability. They introduce a failure probability into the target coverage problem to improve and control the system reliability. They modeled the solution as α-Reliable Maximum Sensor Covers (α-RMSC) problem and proposed a heuristic greedy method to find maximum number of α-Reliable sensor covers and their algorithm can control the failure rate of whole system which a critical aspect in many applications of wireless sensor networks such as military surveillance systems, and environment monitoring systems.

In [32], authors devised a polynomial-time constant-approximation for *M*inimum *W*eight *S*ensor *C*overage *P*roblem (MWSCP). They proposed a polynomial-time $(4 + \varepsilon)$-approximation algorithm for MWSCP. A learning automaton based algorithm to find maximum disjoint set covers of target coverage proposed in [30]. They used from learning automata to schedule node into disjoint set cover to monitor all targets in network. Mostafaei developed an Imperialist Competitive

Algorithm (ICA) based approach to extend the network's lifetime [33]. In this work, author used from characteristics of ICA to find best nodes in each time to monitor deployed targets in network.

## 3. PROBLEM STATEMENT

The maximum lifetime coverage problem in wireless sensor networks formally define as follow: given a sensor network of N sensor nodes and T targets which are randomly deployed within a $L \times L$ rectangular area $\Omega$. Suppose that S be a set of sensor nodes $\{S_1, S_2, ..., S_n\}$ and T be a set of targets $\{t_1, t_2, ..., t_m\}$ with location information and assume that all sensor nodes in network has equal sensing radius and can switch between active and sleep modes. Also, we suppose that the number of sensor nodes that deployed in monitored area is greater than it is required for monitoring target information. We like to schedule the activity state of the sensor nodes to save their energies and improve the network lifetime.

In the proposed method, the following notation is taken;

- A sensor network of N sensor nodes
- T fixed targets which are randomly deployed
- A $L \times L$ rectangular area $\Omega$.
- S be a set of sensor nodes $\{S_1, S_2, ..., S_n\}$
- M the number of targets.
- T be a set of targets
- $W_i$ the lifetime of sensor $S_i$
- $t_m$ the $m$th target, $1 \leq m \leq M$.
- $S_i$ the $i$th sensor, $1 \leq i \leq N$.
- $\lambda$ the time that each set cover is active

If the Euclidean distance between a sensor node and a target is less than the sensing radius of a node, the node can monitor this target. The covered target list of a sensor node $s_i$ is defined as the list of the targets $s_i$ can monitor.

The main problem here is how to organize sensor into several cover sets in which each cover set could monitor all the targets and, at the same time, the network lifetime could be maximized. In this paper, organizing the sensors refers to specifying the mode of the sensors as either active or passive.

Theorem 1: Maximum Set Cover problem is NP-complete [4].

## 4. LEARNING AUTOMATA

A learning automaton is an adaptive decision-making tool that operates in unknown random environments. It has a finite set of actions to choose at each state and choose an action based on action probability vector. For each action that chosen by learning automaton, the environment gives a signal based on probability distribution. The automaton update its action probability based on reinforcement signal that environment gives to random selected action. The main objective of learning automaton is to find optimal action among action set. It tries to minimize average penalty that received from the environment. Figure 1 illustrates relationship between automaton and environment.

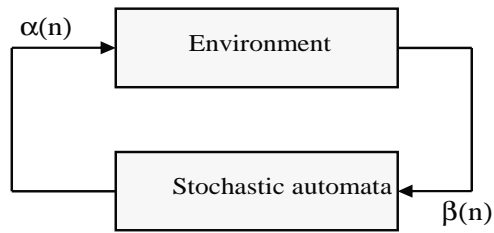

Fig 1 The relationship between the learning automaton and its random environment.

Environment can be described by the triple $E \equiv \{\alpha, \beta, c\}$ where $\alpha \equiv \{\alpha_1, \alpha_2, ..., \alpha_r\}$ denotes finite input set, $\beta \equiv \{\beta_1, \beta_2, ..., \beta_r\}$ represents the output set that can be given by reinforcement signals, and $c \equiv \{c_1, c_2, ..., c_r\}$ is a set of penalty probabilities, where each element $c_i$ of c corresponds to one input of action $\alpha_i$. The environment can be classified into: P-model, Q-model, and S-model based reinforcement signal. The environment in which β can take only two binary values 0 or 1 is referred to P-model environment. Another class of the environment allows a finite number of the values in the interval [0, 1] can be taken by the reinforcement signal. Such an environment is referred to as Q-model environment. In S-model environments, the reinforcement signal lies in the interval [a, b]. Learning automata are classified into fixed-structure stochastic, and variable-structure stochastic. In the following, we consider only variable-structure automata [10].

A learning algorithm can be defined as follows $p(n+1) = T[p(n), \alpha(n), \beta(n)]$. Let $\alpha(k)$ and p(k) denote the action chosen at instant k and the action probability vector on which the chosen action is based, respectively. The repetition equation shown by (1) and (2) is a linear learning algorithm by which the action probability vector p is updated. Let $\alpha_1(k)$ be the action chosen by the automaton at instant k.

$$p_i(n+1) = p_i(n) + a[1 - p_i(n)]$$
$$p_j(n+1) = (1-a)p_j(n) \qquad \forall j, \; j \neq i \tag{1}$$

when the taken action is rewarded by the environment (i.e., $\beta(n) = 0$) and

$$p_i(n+1) = (1-b)p_i(n)$$
$$p_j(n+1) = \frac{b}{r-1} + (1-b)p_j(n) \qquad \forall j, \; j \neq i \tag{2}$$

when the selected action is penalized by the environment (i.e., $\beta(n) = 1$), r is the number of actions that can be a and b denote the reward and penalty parameters and determine the amount of increases and decreases of the action probabilities, respectively. In these two equations, *a* and *b* are reward and penalty parameters respectively. For $a = b$, learning algorithm is called Linear Reward-Inaction ($L_{R-I}$) algorithm, for $b << a$, it is called Linear Reward epsilon Penalty ($L_{R-\varepsilon P}$) algorithm, and for $b = 0$, it is called linear reward–penalty ($L_{R-P}$) algorithm. In [13, 14, 27, 28, 30, 31] some usage of learning automata for wireless sensor networks are introduced.

## 4.1 Action Probability Updating

In this section we restate the *variable structure stochastic automata* (VSSA). VSSA are the ones in which the state transition probabilities are not fixed. In such automata, the state transitions or the action probabilities themselves are updated at every time instant using a suitable scheme. The transition probabilities and the output function in the corresponding Markov chain vary with time, and the action probabilities are updated on the basis of the input. VSSA depend on random number generators for their implementation. The action chosen is dependent on the action probability distribution vector, which is, in turn, updated based on the reward/penalty input that the automaton A variable-structure automaton is defined by the quadruple $\{\alpha, \beta, P, T\}$ in which $\alpha = \{\alpha_1, ..., \alpha_n\}$ represents the action set of the automata, $\beta = \{\beta_1, ..., \beta_n\}$ represents the input set, $P = \{P_1, ..., P_n\}$ represents the action probability set, and finally $p(n+1) = T[\alpha(n), \beta(n), p(n)]$ represents the learning algorithm. This automaton operates as follows. Based on the action probability set $p$, automaton randomly selects an action $\alpha_i$, and performs it on the environment. After receiving the environment's reinforcement signal, automaton updates its action probability set based on equations (1) for favorable responses, and equations (2) for unfavorable ones.

## 5. PROPOSED ALGORITHM

In this section, we describe our proposed method. Network operations divided into different rounds. Each round starts initial phase and continues with learning phase, and ends with target monitoring phase. In the initial phase all nodes of the network participate. At the end of this phase, all sensor nodes in network are aware from their neighbors and monitored targets. In learning phase which is performed periodically during the normal operation of the network, each node with the help learning automata learns to be either active or idle during current round. Finally, in the target monitoring phase, each node selects best actions based of learned information which is active or idle. In our approach, we have two type of packet; *INITIALIZATION* packet and *LAP* packet. In first phased we used *INITIALIZATION* packet to identify nodes neighbors and covered target list of each node and also we used *LAP* packet during learning phase to give reward and penalty for selected action of each node in any time.

### 5.1 Initial Phase

In this phase we equip each node in network with a learning automaton. Learning automata of each node has two actions; ACTIVE and IDLE. At the beginning of the algorithm, ACTIVE and IDLE actions have the same probability equal to 0.

First, each node senses its surrounding environment and determines its covered target list. Then, each node broadcasts an *INITIALIZATION* packet in its neighborhood, containing its ID, position and covered target list. The node then listens to receive *INITIALIZATION* packets from its neighbors. From here on, the network operation is divided into a number of rounds. Each round begins with a learning phase, followed by a target monitoring phase.

### 5.2 Learning Phase

During the learning phase, each node in network performs as follows: we select a random node in network and learning automata of this node randomly selects one of its actions and create a LAP packet. After creating LAP packet, this node puts its status in packet and broadcast it to all neighbors' nodes that have in its neighbor list. Each neighbor node selects one of its actions based on learning automata and sends it to sender node. When sender node received all reply from its neighbors act as follow; if the selected action of *LA by this node* was ACTIVE then If all of the targets under the coverage of the node are covered (not covered) by those neighbors whose selected actions are ACTIVE, then node penalizes (rewards) its selected action and vice versus. This process will continue until all targets in network covered. We do this while the end of learning phase con-dition occurs. We used from action probability criteria to pass learning phase and we supposed that one actions probability passes 0.85.

**5.3 Target Monitoring Phase**

At the end of the learning phase, and at the beginning of a new target monitoring phase, each node selects its state for the whole duration of the current monitoring phase according to the action probability vector of its learning automaton. If the action probability of ACTIVE action is higher than 0.85, the state of this node will be active and vice versus. An active node will monitor the targets in its sensing range for the whole duration of the target monitoring phase. A sleep node does nothing and just saves its battery for future rounds.

*Definition 1:* Duration of Target Monitoring Phase

We suppose each node in target monitoring phase monitors the targets for $\psi$ units of time.

Figure 2 demonstrates the pseudo code of proposed learning automata based method to maximize network lifetime in wireless sensor network.

```
The LAML algorithm
Input:
    (i) Given a set S of N sensor node
    (ii) A set T of M targets and sensing range.
    (iii) iters=total number of iterations
    (iv) α, β learning parameter
Output:
    A converged network`s targets that has monitor all targets
BEGIN
    Do Initial Phase
        While(All Targets Can Cover With sensor nodes)
            Do Learning Phase
            Do TARGET MONITORING PHASE
        End While
END
```

Fig 2 pseudo code of proposed learning automata

## 6. SIMULATION RESULTS

In this section, we conduct a set of simulations to evaluate the performance of the proposed scheduling mechanism, referred to as LAML, in comparison to the performance of similar existing method. All the experiments are implemented in C# and run on a core i5 CPU 2.5-GHz machine with 3-G RAM. In these simulations, a fixed sensor network is assumed, in which all sensor nodes are randomly scattered throughout a 500m × 500m two dimensional area. A number of fixed targets are also deployed randomly within this area. Sensing ranges of all sensor nodes assumed to be equal. Parameters of the conducted simulations are as follows; N: Number of sensor nodes. We vary *n* in the range [20, 80] to study the effect of the node density on the performance of LAML. T: Number of targets. We vary m in the range [10, 50]. R: Sensing range of the sensor nodes. We vary R in the range [100, 600] meters and $\psi$ set to 0.2.

We used the first order energy consumption model, given in [7], for estimating the amount of energy consumed for transmission of the packets between sensor nodes in the network. Energy required to switch a node from sleep to active mode is assumed to be negligible. Results are averaged over 50 runs.

## 6.1 The impact of learning automata on network lifetime

In this section, we first study how much longer lifetime we can achieve by increasing nodes number. Figure 3a shows for 20 sensors and 15 targets, increasing the sensing range results in increasing network lifetime. In this experiment, the lifetime is not sensitive to the number of targets and with doubling the number of targets the network`s lifetime decrease rarely. Figure 3b shows for 30 targets and sensing range 300, increasing the number of sensors will get more network lifetime in our method. When the sensing range decreases to 250, the network`s lifetime considerably go down.

For large networks, we apply our learning automata based method to increase lifetime. Obviously, with large networks we can get the same trend as in small networks, in proportion as we increase the number of sensors per targets, the lifetime increase. We set the number of the sensors to 40 sensor nodes and sensing ranges vary between 100 and 500 meter to study the effect of the distribution of sensor nodes with different sensing ranges on the performance of the proposed algorithm. As we can see in the figure 4a, with increasing the sensing range, longer lifetime gaining. We compared Figs. 3a with 4b and observed that even the network lifetime of each curves are very close to each other.

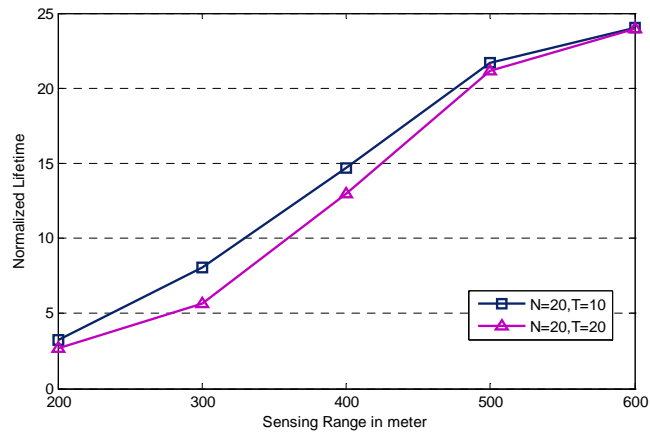

a)

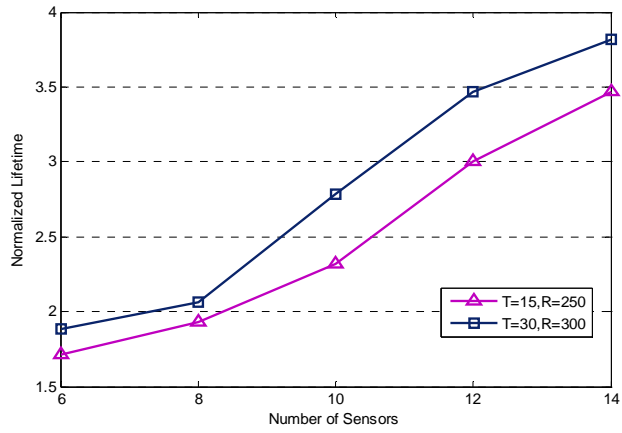

b)

Fig 3. Increasing sensing range from 200 to 600 with $N = 20$, $T = 10$ and 20, respectively; **b** Deploying more sensors, with $N = 6–14$, $T = 15$, $R = 300$ and $T = 30$, range $R = 250$, respectively

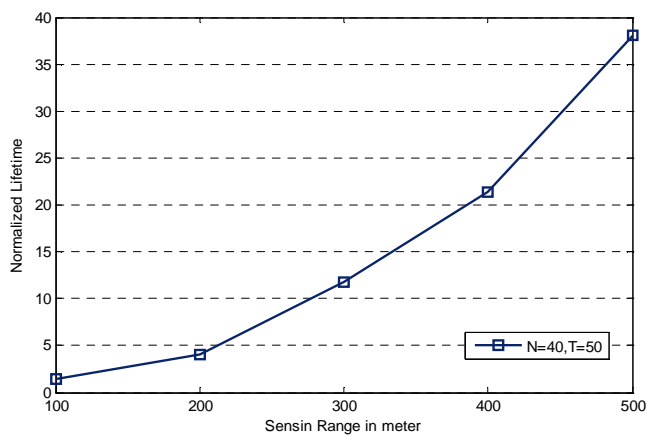

a)

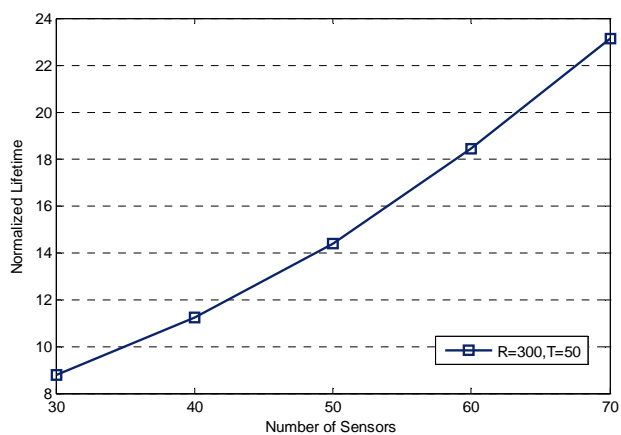

b)

Fig 4. Increasing sensing range from 100 to 500 with 40 sensors and 50 targets; **b** Varying network size from 30 to 70 sensors with fixed range $R = 300$, $T = 50$

## 6.2 LAML versus previous work

Next, we compare our learning automata based algorithm that labeled as LAML with existing work (heuristic Greedy-MSC method) in [4]. For this experiment, we set the number of targets to 50; let the sensing range vary in the range 200 to 500 step by 50, and the number of sensor nodes to 40 to study the effect of the distribution of sensor nodes with different sensing ranges on the performance of the proposed algorithm. We study the effect of the sensing ranges of the sensor nodes on the lifetime of the network in the proposed scheduling mechanism with different sensing ranges. Figure 5a gives the results of this experiment. It can be seen from this figure that the network lifetime is significantly higher when the proposed scheduling mechanism is used rather than heuristic Greedy-MSC method. Next we study the effect of the number of sensor nodes on the lifetime of the network in the proposed scheduling mechanism. Figure 5b shows for sensing range $R = 300$, $N = 20–80$, and $M = 50$. The results of this experiment, which are given in figure 5b, indicate that the network lifetime increases as the number of the sensor nodes increase.

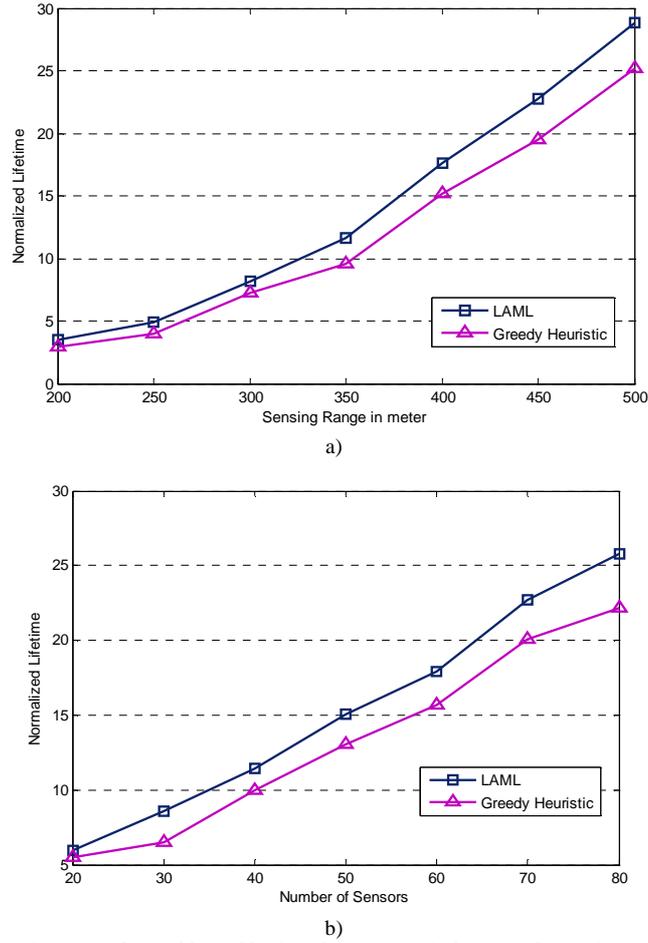

Fig 5. Increasing sensing range from 200 to 500 with 40 sensors and 50 targets; **b** Varying network size from 20 to 80 sensors with fixed range $R = 300$, $T = 50$

**6.3 Impact of learning rate**

In this experiment, we study the impact of the learning rate, used in the proposed algorithm, on the network lifetime. To this end, we consider the following learning rates: 0.01, 0.1, 0.2, and 0.4. Additional simulation parameters are as follows: sensing range is set to 250(m), numbers of deployed targets set, and the network size is set 25 to 50. The result of this experiment, which is given in figure 6, shows that by decreasing the learning rate, the network lifetime also increases. In other words, increasing the (computational and communicational) complexity of the learning phase of the proposed algorithm (by decreasing the learning rate) is not a waste of resources, since this results in more better scheduling of the activity states of the sensor nodes, which consequently results in the network lifetime to increase.

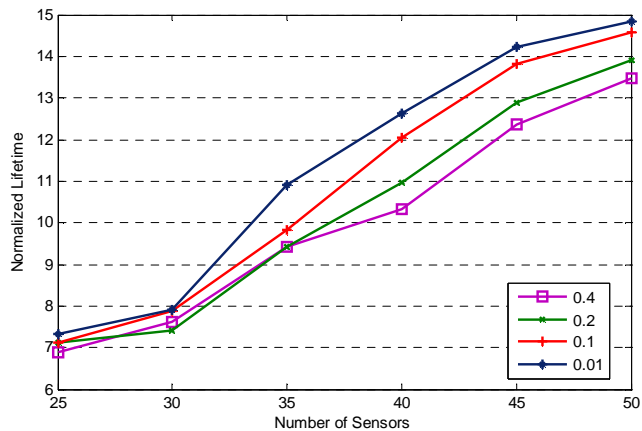

Fig 6. Increasing sensing learning rate from 0.01 to 0.4.

## 7. CONCLUSION

In this paper, we proposed a learning automata-based algorithm for maximum set cover problem in wireless sensor networks. In the proposed algorithm, each node in the network is equipped with a learning automaton. Learning automaton of each node, in cooperation with the learning automata of the neighboring nodes, helps the node to decide its proper activity state to obtain high target coverage. Experimental results showed that the proposed algorithm, regardless of the sensor nodes' density, number of the sensor nodes, and sensing radius of the sensor nodes, outperforms the similar existing methods in terms of the network lifetime.